 \documentclass[final,5p,times,twocolumn,authoryear]{elsarticle}

\usepackage{graphicx} 
\usepackage[english]{babel}
\usepackage{latexsym}
\usepackage{amssymb}
\usepackage{amsmath}
\usepackage{float}
\usepackage{mismath}
\usepackage{xcolor}
\usepackage{natbib}
\usepackage{hyperref}
\usepackage{subcaption}

\begin{document}

\begin{frontmatter}



\title{Domain Wall formation from $Z_2$ spontaneous symmetry breaking/restoration in Scalar-Einstein-Gauss-Bonnet theory}


\author[first,second]{M.A.~Krasnov}
\affiliation[first]{organization={National Research Nuclear University MEPhI},
            city={Moscow},
            postcode={115409}, 
            country={Russia}}
\affiliation[second]{organization={Research Institute of Physics, Southern Federal University},
            city={Rostov-on-Don},
            postcode={344090}, 
            country={Russia}}

\author[third]{D.Z. Berkimbayev}    
\affiliation[third]{organization={Al-Farabi Kazakh National University},
            city={Almaty},
            postcode={050040}, 
            country={Kazakhstan}}

\author[fourth,fifth]{A.~Addazi}
\affiliation[fourth]{organization={Center for Theoretical Physics, College of Physics Science and Technology, Sichuan University},
            city={Chengdu},
            postcode={610065}, 
            country={China}}
\affiliation[fifth]{organization={Laboratori Nazionali di Frascati INFN},
            city={Rome},
            postcode={344090}, 
            country={Italy}}
\author[sixth]{Y.~Aldabergenov}
\affiliation[sixth]{
            organization={Department of Physics, Fudan University}, 
            city={Shanghai},
            postcode={200433}, 
            country={China}}
\author[seventh]{M.~Khlopov}
\affiliation[seventh]{
            organization={Virtual Institute of Astroparticle physics}, 
            city={Paris},
            postcode={75018}, 
            country={France}}

\begin{abstract}
This study offers a detailed analysis of domain wall formation and its cosmological consequences in Einstein-Gauss-Bonnet gravity coupled to a scalar field. A central aspect of the model is the scalar field Lagrangian's ability to spontaneously break and restore its $Z_2$ discrete symmetry.  This spontaneous symmetry breaking is a fundamental prerequisite for topological defect formation. In this context, domain walls arise as kink-like, solitonic solutions that interpolate between the distinct vacuum states of the theory. We perform a detailed numerical analysis of the dynamics of a neutral scalar field non-minimally coupled to the Gauss-Bonnet invariant, exploring its behavior across different cosmological backgrounds. Our results show that coupling to the Gauss-Bonnet term enables the formation of static domain walls with a fixed proper distance within a de Sitter (inflationary) background. Furthermore, we extend our analysis to a radiation-dominated epoch, where we identify that the cosmic expansion causes the "melting" of these domain walls. To assess the potential observational signatures of this scenario, we calculate the predicted spectrum of stochastic gravitational waves generated by the network dynamics using {\it CosmoLattice} package. We also examine the possible generation of Primordial Black Holes (PBHs) associated with collapsing domain walls. Regrettably, 
our calculations indicate that the direct observational detection of such domain walls from this model lies beyond the reach of foreseeable experiments.
Our results constitute a { No-Go} argument against the generation of PBHs as well as of large amplitude GW signals from domain walls in a Scalar-EGB spontaneous symmetry breaking mechanism. 
\end{abstract}




\end{frontmatter}




\section{Introduction}
General Relativity (GR), elegantly encapsulated in the Einstein-Hilbert action, has stood for over a century as the definitive classical description of gravity. Its predictions, from the perihelion precession of Mercury to the recent observations of Gravitational Waves (GWs) \cite{Abbott_2016, Abbott_2017}, have been resoundingly confirmed. Nevertheless, GR is not a complete theory. It predicts its own breakdown in singularities, it is fundamentally classical, and it offers no fundamental explanation for the dark energy driving cosmic acceleration. These shortcomings motivate the search for modified theories of gravity. Among the most well-motivated extensions is Einstein-Gauss-Bonnet (EGB) gravity \cite{YOUSAF2025101221, Yogesh_2025, Dadhich:2005mw}, which incorporates a specific quadratic curvature term that naturally arises in the low-energy effective action of string theory and avoids pathological ghost instabilities \cite{Nojiri_2019, Himmetoglu_2009, Clough_2022}. See e.g. \cite{Fernandes_2022} and references therein for review of EGB gravity. EGB modification of gravity provides a rich phenomenology \cite{Sotiriou_2014, Doneva_2018, Silva_2018, Kleihaus_2011, Odintsov_2020, universe6110197, Gregory_2011, Ghosh_2020}. In addition, this particular extension of GR is capable of explaining accelerated cosmological inflation \cite{Pinto2025, Kanti_2015, ZAHOOR2026100458} and even play a crucial role in baryogenesis \cite{Liang:2019fkj}. Inflation coupled to GB were also considered in \cite{Kanti:2015pda,Jiang:2013gza,Hwang:1999gf,Cartier:2001is,Hwang:2005hb,Guo:2009uk,Guo:2010jr,Koh:2014bka,Kawai:1998ab,Kawai:2017kqt,Yi:2018gse,Rashidi:2020wwg,Kawai:2021bye,Kawai:2021edk,Kawaguchi:2022nku,Addazi:2024gew}.

Our aim is to explore the recently demonstrated Gauss-Bonnet symmetry breaking \cite{Aldabergenov_2025} as a novel mechanism for domain walls formation and study their potential consequences. 

It is known that, in a class of modified gravity models, domain walls, and the BHs formed from their collapse, can arise even without the direct inclusion of matter fields \cite{10.3389/fspas.2022.927144}.
Possibility of formation of topological defects during inflation was studied in \cite{PhysRevD.44.340}. Moreover, topological defects could be formed via quantum tunneling processes mediated by instantons. The process of evolution of domain walls in de Sitter background was studied in several previous works \cite{Basu_1994,Dolgov_2016,Dolgov_2018,Guerrero_2006,NaokiSasakura_2002}. 

In this paper, in the framework of EGB gravity, we introduce a Gauss-Bonnet curvature term coupled with a neutral scalar field with polynomial potential. This work establishes the viability of domain wall formation during inflation and examines the evolutionary dynamics of walls generated in the post-inflationary era. Domain walls produced after inflation 'melt' as their tension decays with cosmological time. Similar cases of melting walls were considered in \cite{Dankovsky_2024, Dankovsky_2025,dankovsky2025cosmicdomainwallslattice,Babichev:2023pbf}. In contrast to aforementioned previous results, we identify and demonstrate a specific physical mechanism responsible for melting domain wall formation. Furthermore, in our model the wall tension exhibits a significantly stronger time dependence, which produces a distinct imprint on the GW spectrum generated by these walls. 
In particular, we establish a no-go result, showing that our model cannot generate significant signals of either GWs or PBHs from domain walls.
Broadly speaking, we show that the domain walls predicted by our model are effectively invisible, leaving almost no observable signatures in the late Universe.

We also utilize {\it  CosmoLattice} package \cite{Figueroa_2021, Figueroa_2023} to study gravitational waves emitted by these melting domain walls.

The paper is organized as follows: Section II is dedicated to the clarification of considered field theory model, in section III we study domain walls which could take place during cosmological inflation and derive constraints on model's parameters in such a scenario, in section IV we describe our setup for lattice simulations via {\it  CosmoLattice} package, in section V we demonstrate gravitational waves spectrum and derive its dependence on initial parameters of the model, then we finalize with discussion and conclusion.



\section{Properties of the model}
In this section we will present the EGB gravity model with neutral scalar field, which could provide a possibility of formation of domain walls.

We consider the model described by the following Lagrangian:
\begin{equation}\label{action}
    \mathcal{L}=\sqrt{-g}\left( \cfrac{R}{2}+g^{\mu\nu}\partial_\mu\phi\partial_\nu\phi-V(\phi)-\cfrac{1}{8}\xi(\phi)R^2_{GB}\right)+ \mathcal{L}_{inf}+\mathcal{L}_{m},
\end{equation}
where Gauss-Bonnet term $R^2_{GB}$ is as follows:
\begin{equation}
    R^2_{GB}=R^2-4R_{\mu\nu}R^{\mu\nu}+R_{\mu\nu\rho\sigma}R^{\mu\nu\rho\sigma}
\end{equation}
and $V(\phi)$, $\xi(\phi)$ are given by
\begin{equation}\label{xi_and_V}
    \xi(\phi)=\cfrac{1}{2}\alpha \phi^2,\, V(\phi)=\cfrac{1}{2}m^2\phi^2+\cfrac{1}{4}\lambda\phi^4.
\end{equation}
We assume $\lambda$ to be positive constant, then preservation of $Z_2$ symmetry is defined by relation between quadratic terms with their constants $\alpha$ and $m$. We assume minimal coupling between scalar field $\phi$ and inflaton field given by  Lagrangian $\mathcal{L}_{inf}$ as well as with any other matter fields represented by $\mathcal{L}_{m}$.

We also assume Friedmann (FLRW) background, which leads to the following equation of motion of a scalar field:
\begin{equation}\label{EOM}
    \Ddot{\phi}+3H\Dot{\phi}-\nabla^2\phi+V_\phi+3\xi_\phi H^2(\Dot{H}+H^2)=0,
\end{equation}
\textcolor{black}{where subscript $\phi$ represents derivative with respect to scalar field $\phi$, whereas dot represents partial derivative with respect to time.}
The presence of Gauss-Bonnet coupling leads to effective potential of the scalar field to be as follows:
\begin{equation}
    V_{eff}(\phi)=V(\phi)+\cfrac{1}{8}\xi(\phi)R^2_{GB}.
\end{equation}
In order to have a possibility of stable domain wall formation we need effective potential to possess two equivalent minima, which could be achieved by choosing $\xi(\phi)$, namely, by choosing the value and sign of the coupling constant $\alpha$. It is of no interest to consider the case of $m^2<0$, since this would lead to the well studied in literature case of domain wall production and Gauss-Bonnet term would simply be (qualitatively) irrelevant. 

\textcolor{black}{We would like to add that there is also a possibility to consider non-minimal coupling between scalar field $\phi$ and gravity in a different form:
\begin{equation}\label{RicciCoupling}
    \mathcal{L}_{int}=\alpha \phi^2 R,
\end{equation}
which is more promising, since scalar curvature does not decrease as fast as GB scalar does after cosmological inflation. However, we emphasize that analysis presented in this paper is dedicated to the study of symmetry breaking induced by GB, which was shown in \cite{Aldabergenov_2025}. Consideration of interaction term given in \eqref{RicciCoupling} would be the subject of our future studies.}

\section{Gauss-Bonnet induced domain walls at de Sitter background}
In this section we study the possibility of formation of cosmic domain walls in de Sitter background. Such background could be referred to cosmological inflation. We assume that inflaton field or whatever causes background to be de Sitter is an external process, where considered scalar field coupled with Gauss-Bonnet term is subdominant effect.

In works by Vilenkin and Dolgov \cite{Basu_1994, Dolgov_2016, Dolgov_2018} planar domain walls were studied. In particular, they derived constraints on model parameters, which defines the possibility of stationary wall's structure in terms of proper distance.

Let us consider Eq.\eqref{EOM} in de Sitter background assuming spherical symmetry \textcolor{black}{(subscript $r$ represents partial derivative with respect to radial coordinate)}:
\begin{equation}
    \Ddot{\phi}+3H\Dot{\phi}-e^{-2Ht}\cfrac{2\phi_r}{r}-e^{-2Ht}\phi_{rr}+V_\phi+3\xi_\phi H^4=0,
\end{equation}
which is then rewritten as:
\begin{equation}\label{SimpleForm}
    \Ddot{\phi}+3H\Dot{\phi}-e^{-2Ht}\cfrac{2\phi_r}{r}-e^{-2Ht}\phi_{rr}+\lambda\phi(\phi^2-\eta^2)=0,
\end{equation}
where $\eta^2=-\left(\cfrac{3\alpha H^4+m^2}{\lambda}\right)$. To have a real positive $\eta$ it is required that $\alpha<-m^2/3H^4$.

Let us now rewrite Eq.\eqref{SimpleForm}, using the following substitutions:
\begin{equation}
\label{substitutions}
    u=Hre^{Ht},\,\phi(u)=f(u)\eta.
\end{equation}
Then we obtain:
\begin{equation}\label{FullEqAfterSub}
    f''(1-u^2)-f'\left(4u-\cfrac{2}{u}\right)=-\cfrac{3\alpha H^4+m^2}{H^2}f(f^2-1).
\end{equation}

We can rewrite it as follows:
\begin{equation}\label{FullEqAfterSubC}
    f''(1-u^2)-f'\left(4u-\cfrac{2}{u}\right)=2Cf(f^2-1),
\end{equation}
where 
\begin{equation}
    2C=-\cfrac{3\alpha H^4+m^2}{H^2}
\end{equation}
to follow the notations introduced in \cite{Basu_1994, Dolgov_2016, Dolgov_2018}. Note that in aforementioned papers it was found that $C>2$ is required for the possibility of static configuration in terms of proper distance. We solve equation Eq.\eqref{FullEqAfterSubC} via the shooting routine since this cannot be solved as a border conditions problem. We use a nonlinear numerical scheme to deal with the stiffness at the origin of the radial coordinate and at the point $u=1$.

\begin{figure}[ht]
	\begin{center}
\includegraphics[width=0.5\textwidth]{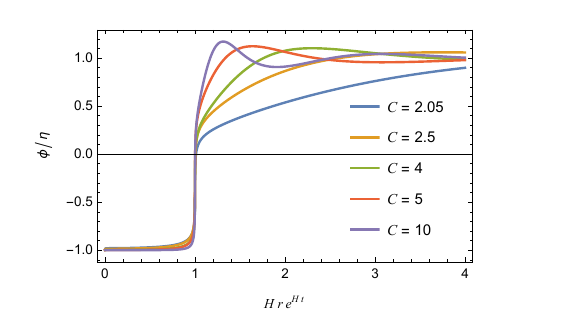}
\end{center}
\vspace{-0.8cm}	
\caption{Numerical solution of Eq.\eqref{FullEqAfterSubC} for different values of $C$. We see that walls are smeared by the expansion as $C$ approaches $2$, whereas bigger $C$ corresponds to thin wall. It is also noticeable that transition region is always at the de Sitter horizon.}
	\label{Res1}
\end{figure}

Now we demonstrate a numerical solution of Eq.\eqref{FullEqAfterSubC} in Fig.~\ref{Res1}. 

From our simulations we see that the threshold for $C$ is the same as in the case of planar walls. 

We should note that de Sitter background is the only possibility to form domain walls with constant tension via Gauss-Bonnet term. If considered scenario takes place during cosmological inflation, then they would disappear during subsequent cosmological evolution because of spontaneous symmetry restoration induced, again, by Gauss-Bonnet term. 

Let us consider Gauss-Bonnet term alone in FLRW background:
\begin{equation}
    R_{GB}^2=24H^2(\Dot{H}+H^2).
\end{equation}
It is clear that during cosmological inflation $R_{GB}^2=24H^4$, which is a positive constant. However, it changes its sign as soon as $\Ddot{a}=0$, which would take place at some moment after inflation ends. For that reason, domain walls induced by Gauss-Bonnet symmetry breaking during cosmological inflation would disappear because of subsequent spontaneous $Z_2$-symmetry restoration. 


\subsection{Loop corrections}

\noindent
For the EGB model defined in \eqref{action}, and in the quasi-de Sitter regime where $H\simeq\text{const}$, the dynamics relevant for domain wall formation is encoded in an effective single field description for $\phi$ with a Gauss--Bonnet induced mass shift. In what follows we include the leading quantum effects in this sector by working with the one-loop effective potential in the Coleman--Weinberg (CW) approximation, evaluated in the slowly varying de Sitter background. This captures the dominant radiative corrections to the scalar self-interactions and to the GB-induced effective mass, while keeping the analysis analytically tractable within the EGB framework.

At tree level the scalar sector is described by
\begin{equation}
V(\phi) = \frac12 m^2 \phi^2 + \frac14 \lambda \phi^4
+ \frac32 \alpha H^4 \phi^2,
\end{equation}
where the last term originates from the Gauss--Bonnet coupling. The GB interaction therefore induces an effective
positive mass shift $\Delta m^2_{\rm GB}=\tfrac{3}{2}\alpha H^4$ for $\phi$.

Quantum fluctuations generate radiative corrections which can be 
systematically included via the one-loop Coleman--Weinberg potential
\cite{Coleman_1973},
\begin{equation}
\Delta V_{1}(\phi) = \frac{M^4(\phi)}{64\pi^2}
\left[ \ln\left(\frac{M^2(\phi)}{\mu^2}\right) - \frac32 \right],
\end{equation}
with the field-dependent mass
\begin{equation}
M^2(\phi) \;=\; m^2 + 3\lambda \phi^2 + 3\alpha H^4.
\end{equation}
The renormalization scale $\mu$ is chosen to minimize large logarithms, typically $\mu\sim\max\{|V''(\phi)|^{1/2},H\}$, and the couplings $(m^2,\lambda,\alpha)$ are understood as running with $\mu$
\cite{Sher_1989,Lyth_Stewart_1992,Herranen_2014,Markkanen_2019}. In regions of
field space where $M^{2}(\phi)$ becomes negative the one-loop expression formally develops an
imaginary part signaling an instability. Following standard practice, and in
order to avoid spurious complex contributions in our perturbative treatment,
we consider $\ln M^{2}(\phi)\rightarrow\ln|M^{2}(\phi)|$
\cite{Weinberg_1987,Weinberg_2005,Parker_Toms_1985}, keeping in mind that this
choice effectively encodes the onset of spinodal dynamics which would require
a real time analysis for a fully self-consistent description.

The one-loop effective potential is then
\begin{equation}
V_{\rm eff}(\phi) = V(\phi) + \Delta V_{1}(\phi).
\end{equation}

\noindent
Variation of the effective action yields the corrected field equation,
\begin{align}
\ddot{\phi} + 3H \dot{\phi}
- e^{-2Ht}\left(\phi_{rr}+\frac{2}{r}\phi_r\right)
+  \phi\left(m^2 + \lambda \phi^2 + 3\alpha H^4\right)
&\nonumber\\
+ \frac{3\lambda\phi}{8\pi^2}\,M^2(\phi)
\left[\ln\left(\frac{M^2(\phi)}{\mu^2}\right)-1\right]
&=0.
\end{align}

\noindent
We again introduce the similarity variable and rescaling \eqref{substitutions}
\begin{equation}
u \equiv Hre^{Ht}, 
\qquad \phi(u)=\eta f(u),
\end{equation}
which reduces the problem to an ODE for $f(u)$:
\begin{align}
H^2\Big[(u^2-1)f''+(4u-\frac{2}{u})f'\Big]
+ m^2 f + \lambda \eta^{2} f^{3} + 3\alpha H^{4} f
&\nonumber\\
+ \frac{3\lambda f}{8\pi^{2}}\mathcal{M}^{2}(f)
\left[\ln\left(\frac{\mathcal{M}^{2}(f)}{\mu^{2}}\right)-1\right]&=0,
\end{align}
with
\begin{equation}
\mathcal{M}^{2}(f)\equiv m^{2}+3\lambda\eta^{2} f^{2}+3\alpha H^{4}.    
\end{equation}

\noindent
Choosing the same normalization
\begin{equation}\label{eta_def}
\eta^{2}\;=\;-\frac{3\alpha H^{4}+m^{2}}{\lambda},
\end{equation}
and defining $\mu_{0}^{2}\equiv m^{2}+3\alpha H^{4}$, we have
$\lambda\eta^{2}=-\mu_{0}^{2}$ and
\begin{equation}
\mathcal{M}^{2}(f)=\mu_{0}^{2}\,(1-3f^{2}).
\end{equation}
Therefore, our ODE becomes
\begin{equation}
\begin{aligned}
&(u^{2}-1)f''+\left(4u-\frac{2}{u}\right)f'
+\frac{|\mu_{0}^{2}|}{H^{2}}\,(f^{3}-f)
\\
&\qquad
+\frac{3\lambda\,|\mu_{0}^{2}|}{8\pi^{2}H^{2}}\,
f(3f^{2}-1)
\left[\ln\!\left(\frac{|\mu_{0}^{2}|(3f^{2}-1)}{\mu^{2}}\right)-1\right]=0.
\end{aligned}
\end{equation}
We can rewrite it in the similar way as \eqref{FullEqAfterSubC}
\begin{multline}\label{FullEqAfterSubCCor}
(1-u^{2})f''-(4u-\tfrac{2}{u})f'=
\\
C\,f\,(f^{2}-1)
+\frac{3\lambda C}{8\pi^{2}}
f(3f^{2}-1)
\left[\ln\left(C(3f^{2}-1)\right)-1\right],
\end{multline}
where we introduced the dimensionless parameter
\begin{equation}
C \equiv \frac{|\mu_{0}^{2}|}{H^{2}},
\qquad
\mu\sim H,
\qquad
\ln\!\big(M^{2}\big)\to\ln\!\big(|M^{2}|\big).
\end{equation}

\begin{figure*}[t!]
    \centering
    \begin{subfigure}[t]{0.5\textwidth}
        \includegraphics{pics/SphericalWall_different_C.pdf}
        \caption{Domain wall's configuration without loop corrections. Same as Fig.\ref{Res1}.}
    \end{subfigure}%
    ~ 
    \begin{subfigure}[t]{0.5\textwidth}
        \includegraphics{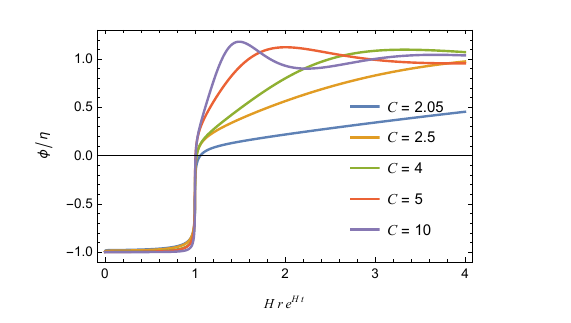}
        \caption{Numerical solution of Eq.~\eqref{FullEqAfterSubCCor} for different values of $C=|\mu_{0}^{2}|/H^{2}$. Here $\lambda=10^{-5}$. Increasing $C$ generally narrows the wall, while the one-loop Coleman--Weinberg term induces a small, parameter-dependent shift of the profile.} \label{Res2}
    \end{subfigure}
    \caption{Comparison of domain wall's static configurations with and without loop correction.}
\end{figure*}

\medskip
\noindent
The GB term shifts the effective mass and together with Hubble
friction create effect against the quartic self-interaction. Equation
\eqref{FullEqAfterSubCCor} therefore describes a ``thick-wall'' profile sourced
by a cubic--quintic nonlinearity and a logarithmic radiative force. The
parameter $C=|\mu_{0}^{2}|/H^{2}$ controls the wall thickness and the relative
weight of mass versus friction: larger $C$ typically sharpens the transition.
The CW contribution modifies the profile slightly and can either sharpen or
smooth the wall depending on $\lambda$ and the local sign of $3f^{2}-1$.
For the numerical solution we impose regularity at the origin,
$f'(u\to0)=0$, and approach to a vacuum at large $u$.  
A fully self-consistent extension could incorporate slow time-dependence of $H(t)$
and RG-improved couplings along the profile, but the quasi-de Sitter and fixed-$\mu$
approximations adopted here capture the leading corrections and are standard in
loop-corrected scalar dynamics in curved spacetime
\cite{Lyth_Stewart_1992,Guo_Schwarz_2010,Kanti_2015,Herranen_2014,Weinberg_2005,Parker_Toms_1985,Markkanen_2019,Figueroa_2023,Weinberg_1987}.
The resulting numerical solutions with one-loop corrections are shown in
Fig.~\ref{Res2}.

\subsection{Constraints on parameters from inflation}
In this section we derive constraint on $\lambda$. Note that this constraint occurs when the $Z_2$ symmetry is broken during cosmological inflation. 

Let us estimate the energy density of the inflaton field $\psi$ as follows:
\begin{equation}
    \rho_{\text{inf}}= \cfrac{1}{2}\Dot{\psi}^2+V_{\text{inf}}\approx V_{\text{inf}}\sim H^2M_{Pl}^2.
\end{equation}
Planck data \cite{refId0} provide constraint on Hubble parameter during cosmological inflation $H<10^{14}\, \text{GeV}$, so, taking $H\sim 10^{13}\,\text{GeV}$, we got
\begin{equation}
   \rho_{\text{inf}}\sim 10^{64}\,\text{GeV}^4;\,E_{\text{inf}}= \rho_{\text{inf}}V_{hor} \sim M_{Pl}^2H^{-1}\sim 10^{25}\, \text{GeV}. 
\end{equation}

Let us now estimate mass of the wall and its energy density.
In flat space-time we got corresponding characteristic width as follows:
\begin{equation}
    \delta_0 = \sqrt{2}/\sqrt{\lambda}\eta=\sqrt{2}/\sqrt{-(3\alpha H^4+m^2)}.
\end{equation}
Surface energy density then is given by
\begin{equation}
    \sigma=\cfrac{4}{3\sqrt{2}}\eta^3\sqrt{\lambda}=\cfrac{4}{3\sqrt{2}}\cfrac{(-\alpha H^4-m^2)^{3/2}}{\lambda}. 
\end{equation}
Finally, for thin wall (assume $|\alpha| H^4 \sim 10H^2 \gg m^2$) we can estimate mass of the wall as follows:
\begin{equation}
    M_{wall}\sim 4\pi H^{-2}\sigma \sim 10^3 \cfrac{H}{\lambda}\sim \cfrac{10^{16}\,\text{GeV}}{\lambda}.
\end{equation}

For energy density of the wall we got:
\begin{equation}
    T^{00}_{max}\sim\cfrac{10^3 H^4}{\lambda}\sim\cfrac{10^{55}\,\text{GeV}^4}{\lambda}.
\end{equation}

Constraint comes from the following condition:
\begin{equation}
    T^{00}_{max} \ll \rho_{\text{inf}} \rightarrow \lambda\geq10^{-8}.
\end{equation}

From Fig.\ref{Res2} we see that $\lambda=10^{-5}$ provides a small correction to the tree-level solution presented in Fig.\ref{Res1} and there are still a few orders of magnitude in a threshold for $\lambda$. 

\textcolor{black}{\subsection{Effective mass of \texorpdfstring{$\phi$}{} during inflation}
To ensure that isocurvature perturbations due to the $\phi$-field are suppressed, its effective mass must be close to the inflationary Hubble scale. It is indeed the case in our models, as one can show by using the functions $\xi$ and $V$ of Eq. \eqref{xi_and_V} to derive the effective mass-squared (around the symmetry-breaking VEV):
\begin{equation}
    V_{{\rm eff},\phi\phi}(\phi=\eta)=2|\mu_0^2|=2CH^2\geq 4H^2~,
\end{equation}
where $\eta$ is given by \eqref{eta_def}, $\mu_0^2\equiv m^2+3\alpha H^4$ and $C\equiv|\mu^2_0|/H^2$, and $C>2$ is imposed as argued above. Consequently, the effective mass of $\phi$ satisfies $m_{{\rm eff}(\phi)}\geq 2H$, which is sufficient to strongly suppress isocurvature perturbations at CMB scales.
 }

\section{Setting up lattice simulations}
In this section we clarify our setup for lattice simulations using {\it  CosmoLattice} package. 

Following numerical analysis in \cite{Dankovsky_2025}, we are going to study this package to calculate spectrum of gravitational waves (GW) emitted by domain walls.

Here we start with some technical details about lattice:
\begin{equation}
    \delta x=\cfrac{L}{N},\, k_{IR}=\cfrac{2\pi}{L}=\cfrac{2\pi}{N\delta x},\,k_{i,UV}=\cfrac{\pi}{\delta x},\,k_{max}=\sqrt{3}k_{i,UV},
\end{equation}
where $\delta x$ is a grid step, $k_{IR}$ is an infrared momentum cut-off corresponding to the grid step, which also defines physical size $L$ of the lattice, $k_{i,UV}$ is ultraviolet momentum cut-off corresponding to one dimension of the grid and $k_{max}$ is defined as maximal momentum, which is calculated as $k_{max}=\sqrt{k_{x,UV}^2+k_{y,UV}^2+k_{z,UV}^2}$. 

Now let us delve into scalar field dynamics. {\it  CosmoLattice} operates with dimensionless units as follows:
\begin{equation}
    \phi \rightarrow \Tilde{\phi} =  \cfrac{\phi}{f_{\star}},\, dx^i \rightarrow d\Tilde{x}^i= \omega_{\star}dx^i, V \rightarrow \Tilde{V}=\cfrac{V}{f^2_\star \omega^2_\star},
\end{equation}
where $f_{\star}$ and $\omega_{\star}$ are somehow defined through model's parameters. We would specify our values of $f_{\star}$ and $\omega_{\star}$ below. 

Following recommendations given in {\it  CosmoLattice} manual \cite{Figueroa_2023}, we would simulate field's dynamics in conformal FLRW spacetime. This particular recommendation follows from field's potential. Namely, authors suggest to utilize conformal FLRW metric if scalar field's potential contains polynomial terms up to quartic, which is our case. 

We utilize vacuum initial conditions with quantum fluctuations defined by power spectrum. Initial quantum vacuum fluctuations could be written as follows:
\[ \langle \delta \phi^2 \rangle = \int d \log k\, \Delta_{\delta \phi}(k),\,\Delta_{\delta \phi}(k)=\cfrac{k^3}{2\pi^2} \mathcal{P}_{\delta \phi}(k),\, \langle \delta \phi_k \delta \phi_{k'} \rangle\]
\begin{equation}
   = (2\pi)^3\mathcal{P}_{\delta \phi}(k)\delta(\mathbf{k}-\mathbf{k'}),
\end{equation}
where $\langle \dots \rangle$ represents an average value and the power spectrum is given as follows: 
\begin{equation}
    \mathcal{P}_{\delta \phi}(k) = \cfrac{1}{2a^2\omega_{k,\phi}},\,\omega_{k,\phi}=\sqrt{k^2+a^2m^2_\phi},\,\,\,\,m_\phi^2=\cfrac{\partial^2 V}{\partial\phi^2}\Bigg{|}_{\phi=\phi_*},
\end{equation}
where $V$ contains Gauss-Bonnet term.
In this expression, $\omega_{k,\phi}$ is the frequency of the mode, and $m_\phi^2$ is the effective mass of the scalar field,
evaluated by the initial homogeneous amplitude $\phi_*$ of the given scalar field. 

Initial fluctuations are governed by $k_{cut}$, which represent the cut-off in the spectrum of fluctuations. This cut-off is defined by user. 

Now we specify the model to simulate scalar field's dynamics. Let us consider the following Lagrangian:
\begin{equation}\label{SimLagr}
    \mathcal{L}=\sqrt{-g}\left( \cfrac{R}{2}+g^{\mu\nu}\partial_\mu\phi\partial_\nu\phi-V(\phi)-\cfrac{1}{8}\xi(\phi)R^2_{GB}\right),
\end{equation}
where $V$ and $\xi$ are defined in the same way as in the previous section:
\begin{equation}\label{VandXiLattice}
    \xi(\phi)=\cfrac{1}{2}\alpha \phi^2,\, V(\phi)=\cfrac{1}{2}m^2\phi^2+\cfrac{1}{4}\lambda\phi^4.
\end{equation}

In conformal FLRW given by line element
\begin{equation}
    ds^2=a^2(\tau)(d\tau^2-d\mathbf{x}^2),
\end{equation}
GB-term is found to be equal to:
\begin{equation}
    R^2_{GB}=\frac{24 \left(a a'^2 a''-a'^4\right)}{a^8},
\end{equation}
where prime represents derivative with respect to conformal time $\tau$.
We can now write equation of motion for the scalar field in conformal FLRW:
\begin{equation}
    \cfrac{1}{a^2}{\phi''}+2\cfrac{a'}{a^3}{\phi'}-\cfrac{1}{a^2}\partial_i^2\phi+V_\phi+\cfrac{1}{8}\xi_\phi R^2_{GB}=0
\end{equation}
and alternatively:
\begin{equation}\label{EoMconf}
    {\phi''}+2\cfrac{a'}{a}{\phi'}-\partial_i^2\phi+a^2\left(V_\phi+\cfrac{1}{8}\xi_\phi R^2_{GB}\right)=0.
\end{equation}
We are interested in radiation dominated (RD) epoch, then the scale factor is as follows: 
\begin{equation}
    a(\tau)=(\tau/\tau_0),
\end{equation}
where $\tau_0$ refers to the start of field's motion (or start of the simulation).

In the process of simulations we have found that potential energy of the scalar field $\phi$ becomes negligible compared to $\Dot{\phi^2}$ and $(\nabla \phi)^2$ after just one Hubble time. It is easy to understand, because in our model symmetry is initially broken by GB term which decreases rapidly with time. We remind that in conformal FLRW in radiational background it is as follows:
\begin{equation}
    R^2_{GB}=\frac{24 \left(a a'^2 a''-a'^4\right)}{a^8}\propto \tau^{-8}.
\end{equation}

Now let us put Eq.\eqref{VandXiLattice} in Eq.\eqref{EoMconf}, then we obtain
\begin{equation}
    {\phi''}+2\cfrac{a'}{a}{\phi'}-\partial_i^2\phi+\lambda \phi a^2\left(\phi^2+\cfrac{m^2}{\lambda}-3\cfrac{\alpha}{\lambda\tau_0^4(\tau/\tau_0)^8}\right)=0.
\end{equation}
We can now introduce vacuum expectation value (VEV):
\begin{equation}
    \eta^2(\tau)=\cfrac{3\alpha}{\lambda\tau_0^4(\tau/\tau_0)^8}-\cfrac{m^2}{\lambda}.
\end{equation}

Given with definition of VEV, we can now specify $f_{\star}$ and $\omega_{\star}$. In our model we set them to be equal to:
\begin{equation}
    f_\star = \eta(\tau_{i}),\, \omega_\star = \sqrt{\lambda}f_\star.
\end{equation}

Equation \eqref{EoMconf} could be simplified dramatically. Let us substitute 
\begin{equation}
    \phi=\cfrac{s}{a},
\end{equation}
then we obtain:
\begin{equation}
    s''-\partial_i^2s+\lambda s(s^2-a^2\eta^2)-\cfrac{a''}{a}s=0.
\end{equation}
During RD epoch we have $a''=0$, so we have very simple EoM:
\begin{equation}\label{AfterSubstitution}
    s''-\partial_i^2s+\lambda s(s^2-a^2\eta^2)=0.
\end{equation}

Equation \eqref{AfterSubstitution} is basically equation of motion in Minkowski space. Let us introduce $\eta'=a\eta$, then in terms of field $s$ we have the following time dependence of wall's tension on time:
\begin{equation}  \label{tension}\sigma(\tau)=\cfrac{2\sqrt{2\lambda}\eta'^3}{3}=\cfrac{2\sqrt{2\lambda}}{3}\left( \cfrac{3\alpha}{\lambda\tau_0^4(\tau/\tau_0)^6}-\cfrac{m^2(\tau/\tau_0)^2}{\lambda}\right)^{3/2}.
\end{equation}
One can easily found the time when $Z_2$ symmetry is restored, if it is initially broken ($\eta(\tau_0)>0$):
\begin{equation}
    \sigma(\tau_{end})=0 \rightarrow \tau_{end}=\sqrt{\tau_0}\left(\cfrac{3\alpha}{m^2} \right)^{1/8}.
\end{equation}
Given with such time dependence of VEV and wall's tension, it is now clear that walls would melt rapidly within this particular model introduced by Lagrangian \eqref{SimLagr}.

In the following sections we would study possible observable consequences of scalar field's motion coupled with Gauss-Bonnet term. Namely we would study emission of gravitational waves and possibility of black hole formation.

We should also note that {\it  CosmoLattice} by default define fraction of energy density of gravitational waves as $\Omega_{GW}=\cfrac{\Tilde{\rho}_{GW}}{\Tilde{\rho}_{tot}} \neq \cfrac{\Tilde{\rho}_{GW}}{\Tilde{\rho}_{c}}$, where $\rho_{tot}$ refers to the energy density of matter fields which are programmed by user, so it does not take into account critical density of the Universe. For that reason output should be multiplied by corresponding factor to restore common definition.
In program units critical density would be as follows:
\begin{equation}
    \Tilde{\rho}_c(\tau)=\cfrac{\rho_c(\tau)}{f^2_\star \omega^2_\star}=\cfrac{3\lambda M^2_{Pl}}{8\pi H_0^2 a^4(\tau)}
\end{equation}
and standart definition is then restored as follows:
\begin{equation}
    \Omega_{GW}=\cfrac{\Tilde{\rho}_{GW}}{\Tilde{\rho}_{tot}} \cdot \cfrac{\Tilde{\rho}_{tot}}{\Tilde{\rho}_{c}}.
\end{equation}
Throughout our paper we would operate with standart definition of fraction of energy density.

We would like to finalize this section by clarifying exact values, which we utilize in our simulations. We set $N=256, k_{IR}=0.1$ and $k_{cut}=4$. 

\section{GWs emitted during RD epoch}
In this next section we present our results of gravitational wave's spectra assuming background is radiation dominated. We utilize vacuum initial conditions with initial fluctuations defined in the previous section. 

For now, let us specify the value of $\alpha$, which is coupling constant between scalar field and Gauss-Bonnet term. Classical motion of the field starts when its effective mass is of order of Hubble parameter, for that reason we choose
\begin{equation}\label{AlphaRd}
    \alpha=\cfrac{1}{3H^2_0},
\end{equation}
where $H_0$ is the Hubble parameter at the start of the motion (simulations). We should note that in radiational background we need positive value of coupling constant $\alpha$ to have initially broken $Z_2$ symmetry. We would like to remind the reader that we assume $m^2>0$ and we set this term to be small so that it is negligible compared with Gauss-Bonnet coupling term at the start of the motion in order to have a possibility to form walls.

Given with Eq.\eqref{AlphaRd}, we can now represent $f_{\star}$ and $\omega_{\star}$ via Hubble parameter at the start of the motion $H_0$ and $\lambda$:
\begin{equation}
    \eta(\tau_0)=\cfrac{H_0}{\sqrt{\lambda}},\,\omega_{\star}=H_0.
\end{equation}
At this point our model has three free parameters: $H_0$, $m$ and $\lambda$.

Now we study the impact of model's parameters on gravitational wave's spectra. 

\begin{figure}[H]
	\begin{center}
\includegraphics[width=0.5\textwidth]{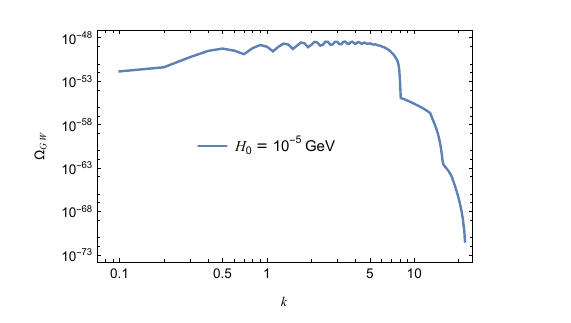}
\end{center}
\vspace{-0.8cm}	
\caption{Spectrum of GWs during RD stage. Duration of simulations $\tau_f = 10\tau_i$ and $m=10^{-3}H_0$.}
\label{GWRD_mass}
\end{figure}

\begin{figure}[ht]
	\begin{center}
\includegraphics[width=0.5\textwidth]{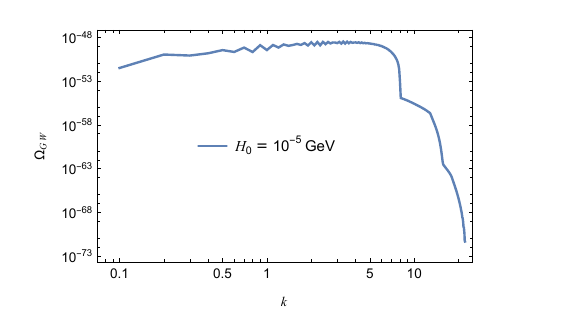}
\end{center}
\vspace{-0.8cm}	
\caption{Spectrum of GWs during RD stage. Time of simulations $\tau_f = 10\tau_i$ and $m=0$.}
	\label{GWRD1}
\end{figure}

Let us study impact of $m$. Presence of non-zero positive mass would eventually lead to spontaneous restoration of $Z_2$ symmetry. We can now calculate a certain moment of time $\tau_{end}$ at which symmetry is restored.  Let us set $m=\beta H_0$, then we obtain:
\begin{equation}
    \tau_{end}=\sqrt{\tau_0}\left(\cfrac{3\alpha}{m^2} \right)^{1/8}=\cfrac{1}{H_0 \beta^{1/4}}.
\end{equation}
Let us assume $\beta = 10^{-3}$, then
\begin{equation}
    \tau_{end}\approx 5.62 \tau_{i}.
\end{equation}
Given with that value above we set the duration of simulations to be $\tau_f =10\tau_0$.  We demonstrate GW spectra for $H_0=10^{-5}\text{GeV}$ and $\lambda=10^{-8}$. Please, see Fig.\ref{GWRD_mass} and Fig.\ref{GWRD1}.

From Fig.\ref{GWRD_mass} and Fig.\ref{GWRD1} one can notice a difference in between plots in the area of small momentum, but this area is far from the peak of the spectrum, thus impact of mass could be observed only if waves corresponding to these momenta could be detected. One may see the value of $\Omega_{GW}$ is extremely low here and cannot be detected by modern detectors. This result seems natural because Gauss-Bonnet term derceases rapidly with time and mass term in the model is set to be small by assumption. Here we conclude that the parameter $m$ is irrelevant in terms of the detectability of the spectrum.

\begin{figure}[ht]
	\begin{center}
\includegraphics[width=0.5\textwidth]{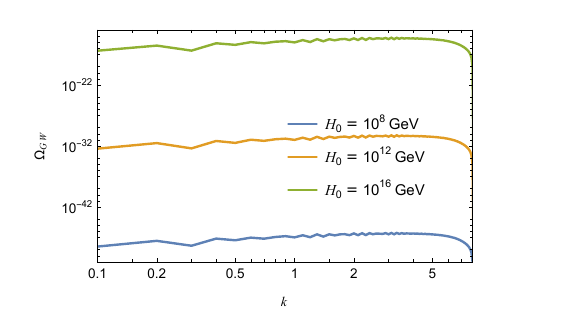} 
\caption{Spectrum of GWs during RD stage. Here we compare spectra obtained with different initial Hubble parameters. One can easily see strong dependence on its value. Although even for unrealisticly  high values of $H_0$ does not produce detectable spectrum.}\label{ImpactOfH}
\end{center}
\end{figure}

Then we progress by fixing value of $\lambda=10^{-8}$ and study impact of initial Hubble parameter. Here we also set $m=0$ to simplify simulations. We demonstrate our findings in Fig.\ref{ImpactOfH}. Gravitational waves are mostly emitted during the start of the filed's motion. By {\it  CosmoLattice} we find that GW energy density starts to behave as a radiational background after just a few Hubble times after the start of the simulations. 

From Fig.\ref{ImpactOfH} we conclude \begin{equation}\label{OmegaDependence}
    \Omega_{GW}(H)=(H/H_0)^4\Omega_{GW}(H_0).
\end{equation}

\begin{figure}[H]
	\begin{center}
\includegraphics[width=0.5\textwidth]{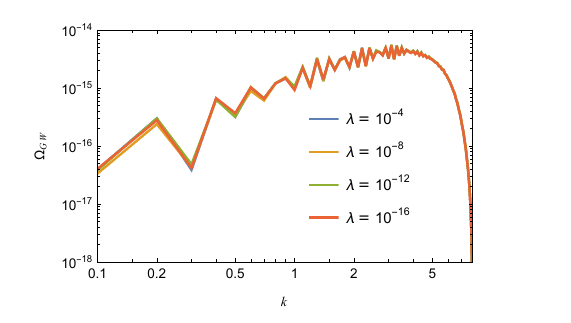} 
\caption{Spectrum of GWs during RD stage. One can see $\Omega_{GW}$ is independent of $\lambda$. Here we set $H_0=10^{16}\text{GeV}$ and $m=0$. }\label{ImpactOfLambda}
\end{center}
\end{figure}

Finally, we study the impact of $\lambda$. We present our findings in Fig.\ref{ImpactOfLambda}.

From the plot in Fig.\ref{ImpactOfLambda} we can see that dependence on $\lambda$ of the spectrum is basically negligible. 

At this point, via ComoLattice simulations we have found that GWs spectra effectively depends only on initial Hubble parameter $H_0$ in our model. 

Now, let us present spectra calculated at modern times as a function of frequency. Here we set the duration of the simulations $\tau_f=20\tau_0$.

Transformation from program variables to frequency is as follows:
\begin{equation}
    F=\cfrac{H_0 k}{2\pi (z_0+1)},
\end{equation}
which means that the frequency depends on the initial Hubble parameter, since $k[GeV]=\omega_\star k = H_0k$. In the expression above $z_0$ is the redshift corresponding to the start of field's motion or simulations.

We now utilize the Hubble law to calculate redshift:
\begin{equation}
    H(z)^2=H_{modern}^2(\Omega_r(1+z)^4+\Omega_m(1+z)^3+\Omega_\Lambda),\, H_{modern}\sim 10^{-42}GeV.
\end{equation}
One can approximately estimate the redshift as:
\begin{equation}
    z_0\approx \left( \cfrac{H(z_0)^2}{\Omega_r H^2_{modern}}\right)^{1/4},
\end{equation}
since we consider RD epoch. By considering different initial Hubble parameters, we are considering different redshifts. 

Finally, frequency corresponding to modern Universe is calculated as follows:
\begin{equation}
    F=\cfrac{(H_0 H_{modern})^{1/2}\Omega_r^{1/4}}{2\pi}k.
\end{equation}

\begin{figure}[ht]
	\begin{center}
\includegraphics[width=0.5\textwidth]{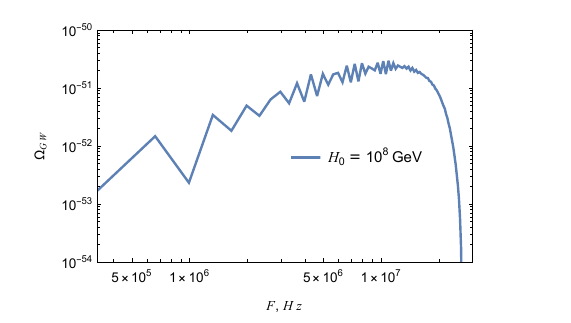}
\end{center}
\vspace{-0.8cm}	
\caption{GWs spectrum as a function of frequency, which corresponds to modern Universe. Here $\Omega_{GW}$ is normalized by modern critical density, taking into account time dependence of GW's energy density.}
	\label{Gws_freq_1}
\end{figure}

\begin{figure}[ht]
	\begin{center}
\includegraphics[width=0.5\textwidth]{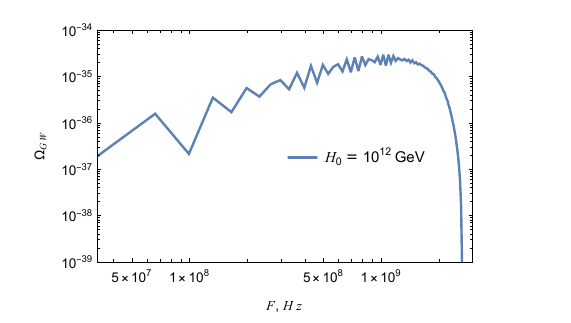}
\end{center}
\vspace{-0.8cm}	
\caption{GWs spectrum as a function of frequency, which corresponds to modern Universe. Here $\Omega_{GW}$ is normalized by modern critical density, taking into account time dependence of GW's energy density.}
	\label{Gws_freq_2}
\end{figure}

We are now ready to present GW spectra corresponding to modern times. Please, see Fig.\ref{Gws_freq_1} and Fig.\ref{Gws_freq_2}.

Let us now estimate strain corresponding to the spectrum in FIg.\ref{Gws_freq_2}. We start  with definition:
\begin{equation}
    \Omega_{GW}=\cfrac{2\pi^2}{3H^2_{modern}}f^3S_h(f)\rightarrow \sqrt{S_h(f)}=\sqrt{\cfrac{3\Omega_{GW}H^2_{modern} }{2\pi^2 f^3}}.
\end{equation}
Taking peak frequency and maximum value of $\Omega_{GW}$ in Fig.~\ref{Gws_freq_2}, estimation of strain is as follows:
\begin{equation}
    \sqrt{S_h}\sim  10^{-48}\sqrt{\text{Hz}}^{-1}. 
\end{equation}

Let us now discuss the possibility of detection of such GWs. From paper \cite{theligoscientificcollaboration2025upperlimitsisotropicgravitationalwave} it is clear that modern facilities are incapable of detecting such weak waves. Is there at least a chance that these waves could be detected by future detectors \cite{cai2025atomicquantumsensorshighfrequency}? In aggressive optical Raman schemes there could be achieved sensitivity up  to
\begin{equation}
    \sqrt{S_{h,min}}\sim 10^{-37}\sqrt{\text{Hz}}^{-1}
\end{equation}
in THz range, which is not applicable to our findings. 

{
Recent researches have demonstrated that in the terahertz range, the graviton-to-photon conversion process can be resonantly amplified by cosmological magnetic fields \cite{Addazi:2024osi,Addazi:2024kbq}. While this makes the detection of very high-frequency gravitational waves via radio astronomy theoretically possible, the signal strength for our specific case remains too weak for any radio-astronomy instruments planned for the foreseeable future.}

Thus, in the future, our model could be found true or constrained. Namely, since we expressed everything in terms of the initial Hubble parameter at the start of the scalar field motion, possible constraints would restrict the value of the coupling constant $\alpha$. 



\textcolor{black}{\subsection{Effective mass of \texorpdfstring{$\phi$}{} during inflation in second scenario}
In this particular scenario we have $Z_2$ symmetry restored during inflation and broken after it. In order to estimate effective mass of the field $\phi$ during inflation, we calculate
\begin{equation}
    V_{\text{eff},\phi\phi}(\phi=0)=m^2+3\alpha H^4_{\text{inf}}=m^2+\cfrac{H^4_{\text{inf}}}{H^2_0}>H^2_{\text{inf}},
\end{equation}
where we have used \eqref{AlphaRd} and $H_{\text{inf}}$ is a Hubble parameter during inflation and $H_0$ is a Hubble parameter at the start of field's motion (physically relevant are $H_0<H_{\text{inf}})$, therefore inflationary isocurvature fluctuations of the field $\phi$ are suppressed in this scenario as well.}

\section{Primordial Black Hole formation: a no-go argument}
In this section we study the possibility of PBH formation within considered model. 

Let us start with clarification of wall's width dependence on the scale factor:
\begin{equation}
    \delta_{wall}(\tau)=\cfrac{\sqrt{2}}{\sqrt{\lambda}\eta'(\tau)}=\cfrac{\sqrt{2}}{\sqrt{\cfrac{3\alpha H_0^4}{a^6}-m^2a^2}}>\cfrac{a^3\sqrt{2}}{H^2_0\sqrt{3\alpha}}=\cfrac{a^3\sqrt{2}}{H_0}=\cfrac{a^3\sqrt{2}}{\omega_{\star}}.
\end{equation}
From the expression above, we conclude that walls in our model are smeared by the expansion, unlike in the case of walls with constant surface energy density.  

We recall tension given by Eq.\eqref{tension}:
\begin{equation}
 \sigma(\tau)=\cfrac{2\sqrt{2\lambda}\eta^3}{3}=\cfrac{2\sqrt{2\lambda}}{3}\left( \cfrac{3\alpha}{\lambda \tau_0^4(\tau/\tau_0)^6}-\cfrac{m^2(\tau/\tau_0)^2}{\lambda}\right)^{3/2}.
\end{equation}
One can estimate the mass of the domain wall $m_w$ as follows:
\begin{equation}
    m_w(\tau)=4\pi \sigma(\tau)u_w^2,
\end{equation}
where $u_w$ is the radius of the wall. This leads to the following relation between the wall's gravitational radius and the wall's radius:
\begin{equation}
    u_g(\tau)=2Gm(\tau)=8\pi G \sigma(\tau)u_w^2(\tau).
\end{equation}
Now let us compare $u_w$ and $u_g$:
\begin{equation}
    \cfrac{u_g}{u_w}=8\pi G \sigma(\tau)u_w(\tau).
\end{equation}
To form a black hole it is required
\begin{equation}\label{ConditionOnU}
     \cfrac{u_g}{u_w}>1\rightarrow u_w(\tau)>\cfrac{3\lambda}{16\pi G \sqrt{2}\left( \cfrac{3\alpha}{\tau_0^4a^6(\tau)}-m^2a^2(\tau)\right)^{3/2}}.
\end{equation}
Expression Eq.\eqref{ConditionOnU} shows the condition under which it is possible to form a PBH. Let us put $m=0$ to weaken that condition:
\begin{equation}
    u_w(\tau)>\cfrac{3\lambda}{16\sqrt{2}\pi G H^3_0}a^9(\tau).
\end{equation}
Expression above indicates that, in order to make PBH formation possible, it is demanded a walls expansion rate which is faster than $a^9$.
Such a possibility is forbidden since it demands an impossible superluminal expansion of domain walls. Walls are considered as being at rest relative to the Hubble expansion,  implying their radius evolves as the scale factor with time $u_w=u_0a$, where $u_0$ is initial radius of the wall. 

Now let us compare gravitational radius to wall's width:
\begin{equation}\label{WidthToG}
    \cfrac{u_g}{\delta}=\cfrac{8\pi G \sigma u^2_w}{\delta}=\cfrac{16\pi G u_0^2 H_0^4}{3\lambda a^{10}} \propto \cfrac{1}{a^{10}},
\end{equation}
which demonstrates even stronger condition compared to wall's radius.

From the expressions above we conclude that it is impossible to form black holes within this model (from the collapse of domain walls). Conditions given by \eqref{ConditionOnU} and \eqref{WidthToG} cannot be satisfied in reality. 

\section{Discussions and Conclusions}
In this paper, we analyzed possible observable effects arising from the spontaneous $Z_2$ symmetry breaking or restoration that is catalyzed by the Gauss-Bonnet term. Our investigation is built upon a specific formulation of the scalar-Gauss-Bonnet coupling constant $\alpha$.

Our investigation commenced with an analysis of domain wall formation in a de Sitter (inflationary) background. Throughout this epoch, the Gauss-Bonnet term remains nearly constant.  Our analysis demonstrated the existence of static domain wall solutions characterized by a constant proper width. The threshold for parameter $C$ in our model was found to be the same as in previous works by {\it Dolgov \& Vilenkin} \cite{Dolgov_2016,Dolgov_2018, Basu_1994}. Finally, we found that in the spherically symmetric case, the domain wall's transitional region is located precisely at the de Sitter horizon. 

Subsequently, we examined the scalar field evolution within a radiation-dominated Universe. Here, the Gauss-Bonnet term diminishes rapidly, which provides a specific mechanism for wall's melting. 
 The melting process was found to be very rapid -- much faster than in the cases considered by  {\it Dankovsky, Ramazanov, Babichev et al} \cite{Dankovsky_2024,Dankovsky_2025,dankovsky2025cosmicdomainwallslattice}.
While prior work assumes the VEV scales inversely with the scale factor, our model ties its evolution to the Gauss-Bonnet term, resulting in accelerated decay. Using the {\it CosmoLattice} package, we analyze this scenario to assess GWs emission from the melting walls and the associated PBH production. Our results showed a strong dependence of the GW spectrum on the initial Hubble parameter when the scalar field evolution begins. Unfortunately, the predicted GW spectra are likely to be undetectable even by next-generation observatories. However, future detectors could potentially place mild constraints on the model parameters. The analysis further reveals that, due to the rapid melting process, the wall’s radius and width must expand with impossible superluminal velocity to eventually become gravitationally self-enclosed. Consequently, PBH formation is ruled out in this scenario. \textcolor{black}{Nontheless, such domain walls can have other effects, such as contributing to inhomogeneities in matter distribution \cite{Moffat:2025whe} or modifying CMB power spectra \cite{Lazanu:2015fua,Venkataratnam:2026orb}, which deserve separate investigations.}

To summarize, our model predicts signals below foreseeable detection thresholds. This result serves as a no-go argument for GB theories of this class, and is robust for any regular coupling function $\xi(\phi)$. \textcolor{black}{In future work, one could consider a complex scalar field with a Mexican hat potential coupled to the Gauss–Bonnet term. Such a setup would give rise to a cosmic string network rather than domain walls. Additionally, the phase (axion) of the complex scalar could acquire a periodic potential via instanton effects, leading to axion domain walls bounded by cosmic strings.}

In the present work, we have not accounted for the potential dynamics of domain-wall melting and decay within the early-universe plasma.
Recent numerical studies of gravitational-wave generation from first-order cosmological phase transitions FOCPT indicate that the dominant signal often originates not from the direct collision of bubble walls, but from the subsequent acoustic (sound-wave) oscillations and magnetohydrodynamic (MHD) turbulence induced in the plasma by the walls’ expansion \cite{Hindmarsh:2013xza} -- this case seems also favored as an explanation of NANOGrav excess from FOCPTs \cite{Addazi:2023jvg} (see also \cite{Addazi:2017nmg,Addazi:2020zcj}. By analogy, a rapid melting or “despairing” of a domain-wall network could likewise transfer substantial energy into the surrounding plasma, thereby exciting sound waves and MHD turbulence capable of sourcing a stochastic gravitational-wave background.
The characteristics of such a signal would depend crucially on the network’s scaling properties at decay, the timescale of melting relative to the Hubble time, and the efficiency with which wall energy is converted into bulk fluid motion. A reliable prediction requires dedicated numerical simulations of a coupled scalar-field and relativistic-fluid system, capturing the decay of the network and the resulting plasma dynamics.
Future work should therefore investigate this possibility through large-scale simulations, analogous to those performed for phase transitions, in order to determine whether plasma-mediated effects could dominate the gravitational-wave signature—or even constitute the primary observable signal—of a transient domain-wall network in the Early Universe.

\textcolor{black}{Our preliminary findings suggest that finite temperature effects on the domain walls are suppressed due to their rapid melting after horizon reentry. The short timescale involved leaves insufficient time for thermalization, implying that the process is inherently non-equilibrium in nature. Although these considerations point toward suppressed thermal effects, a more detailed study is needed for definitive conclusions.}



\section*{Acknowledgments}
We are grateful to V. Nikulin and I. Dankovsky for fruitful discussions and interest in our work. The work of M. Kr. was performed in Southern Federal University with financial support of grant of Russian Science Foundation № 25-07-IF.
AA work is supported by the National Science Foundation of China (NSFC) through the grant No.\ 12350410358; the Talent Scientific Research Program of College of Physics, Sichuan University, Grant No.\ 1082204112427; the Fostering Program in Disciplines Possessing Novel Features for Natural Science of Sichuan University, Grant No.2020SCUNL209 and 1000 Talent program of Sichuan province 2021.

\bibliographystyle{elsarticle-harv} 
\bibliography{references}

@ARTICLE{10.3389/fspas.2022.927144,
    
AUTHOR={Nikulin, Valery V.  and Krasnov, Maxim A.  and Rubin, Sergey G. },
           
TITLE={Compact extra dimensions as the source of primordial black holes},
          
JOURNAL={Frontiers in Astronomy and Space Sciences},
          
VOLUME={Volume 9 - 2022},
  
YEAR={2022},
  
URL={https://www.frontiersin.org/journals/astronomy-and-space-sciences/articles/10.3389/fspas.2022.927144},
  
DOI={10.3389/fspas.2022.927144},
  
ISSN={2296-987X}
  }

@article{Aldabergenov_2025,
   title={Gauss–Bonnet-Induced Symmetry Breaking/Restoration During Inflation},
   volume={11},
   ISSN={2218-1997},
   url={http://dx.doi.org/10.3390/universe11030098},
   DOI={10.3390/universe11030098},
   number={3},
   journal={Universe},
   publisher={MDPI AG},
   author={Aldabergenov, Yermek and Berkimbayev, Daulet},
   year={2025},
   month=mar, pages={98} }

@article{Lazanu:2015fua,
    author = "Lazanu, A. and Martins, C. J. A. P. and Shellard, E. P. S.",
    title = "{Contribution of domain wall networks to the CMB power spectrum}",
    eprint = "1505.03673",
    archivePrefix = "arXiv",
    primaryClass = "astro-ph.CO",
    doi = "10.1016/j.physletb.2015.06.034",
    journal = "Phys. Lett. B",
    volume = "747",
    pages = "426--432",
    year = "2015"
}

@article{Venkataratnam:2026orb,
    author = "Venkataratnam, K. K.",
    title = "{Role of Domain Walls in the Early Universe in the Context of Mode Matching}",
    eprint = "2603.01533",
    archivePrefix = "arXiv",
    primaryClass = "astro-ph.CO",
    month = "3",
    year = "2026"
}

@article{Moffat:2025whe,
    author = "Moffat, John",
    title = "{Void and Density Walls Inhomogeneous Cosmic Web Cosmology}",
    eprint = "2502.20494",
    archivePrefix = "arXiv",
    primaryClass = "astro-ph.CO",
    month = "2",
    year = "2025"
}

@article{NaokiSasakura_2002,
doi = {10.1088/1126-6708/2002/02/026},
url = {https://doi.org/10.1088/1126-6708/2002/02/026},
year = {2002},
month = mar,
publisher = {},
volume = {2002},
number = {02},
pages = {026},
author = {Naoki Sasakura},
title = {A de-Sitter thick domain wall solution by elliptic functions},
journal = {Journal of High Energy Physics}
}

@article{Guerrero_2006,
   title={De Sitter and double irregular domain walls},
   volume={38},
   ISSN={1572-9532},
   url={http://dx.doi.org/10.1007/s10714-006-0297-y},
   DOI={10.1007/s10714-006-0297-y},
   number={5},
   journal={General Relativity and Gravitation},
   publisher={Springer Science and Business Media LLC},
   author={Guerrero, Rommel and Rodriguez, R. Omar and Torrealba, Rafael and Ortiz, R.},
   year={2006},
   month=apr, pages={845–855} }

@article{Babichev:2023pbf,
    author = "Babichev, E. and Gorbunov, D. and Ramazanov, S. and Samanta, R. and Vikman, A.",
    title = "{NANOGrav spectral index {\ensuremath{\gamma}}=3 from melting domain walls}",
    eprint = "2307.04582",
    archivePrefix = "arXiv",
    primaryClass = "hep-ph",
    doi = "10.1103/PhysRevD.108.123529",
    journal = "Phys. Rev. D",
    volume = "108",
    number = "12",
    pages = "123529",
    year = "2023"
}

@article{Clough_2022,
   title={Ghost Instabilities in Self-Interacting Vector Fields: The Problem with Proca Fields},
   volume={129},
   ISSN={1079-7114},
   url={http://dx.doi.org/10.1103/PhysRevLett.129.151102},
   DOI={10.1103/physrevlett.129.151102},
   number={15},
   journal={Physical Review Letters},
   publisher={American Physical Society (APS)},
   author={Clough, Katy and Helfer, Thomas and Witek, Helvi and Berti, Emanuele},
   year={2022},
   month=oct }

@article{Himmetoglu_2009,
   title={Ghost instabilities of cosmological models with vector fields nonminimally coupled to the curvature},
   volume={80},
   ISSN={1550-2368},
   url={http://dx.doi.org/10.1103/PhysRevD.80.123530},
   DOI={10.1103/physrevd.80.123530},
   number={12},
   journal={Physical Review D},
   publisher={American Physical Society (APS)},
   author={Himmetoglu, Burak and Contaldi, Carlo R. and Peloso, Marco},
   year={2009},
   month=dec }

@inproceedings{Dadhich:2005mw,
    author = "Dadhich, Naresh",
    title = "{On the Gauss-Bonnet gravity}",
    booktitle = "{12th Regional Conference on Mathematical Physics}",
    eprint = "hep-th/0509126",
    archivePrefix = "arXiv",
    pages = "331--340",
    month = "9",
    year = "2005"
}

@article{Ghosh_2020,
doi = {10.1088/1361-6382/abc134},
url = {https://doi.org/10.1088/1361-6382/abc134},
year = {2020},
month = nov,
publisher = {IOP Publishing},
volume = {37},
number = {24},
pages = {245008},
author = {Ghosh, Sushant G and Kumar, Rahul},
title = {Generating black holes in 4D Einstein–Gauss–Bonnet gravity},
journal = {Classical and Quantum Gravity}
}

@article{Yogesh_2025,
doi = {10.1088/1475-7516/2025/10/010},
url = {https://doi.org/10.1088/1475-7516/2025/10/010},
year = {2025},
month = oct,
publisher = {IOP Publishing},
volume = {2025},
number = {10},
pages = {010},
author = {Yogesh and Mohammadi, Abolhassan and Wu, Qiang and Zhu, Tao},
title = {Starobinsky like inflation and EGB Gravity in the light of ACT},
journal = {Journal of Cosmology and Astroparticle Physics}
}

@article{Gregory_2011,
doi = {10.1088/1742-6596/283/1/012016},
url = {https://doi.org/10.1088/1742-6596/283/1/012016},
year = {2011},
month = feb,
publisher = {},
volume = {283},
number = {1},
pages = {012016},
author = {Gregory, Ruth},
title = {Holographic Superconductivity with Gauss-Bonnet gravity},
journal = {Journal of Physics: Conference Series}
}

@Article{universe6110197,
AUTHOR = {Kokubu, Takafumi and Harada, Tomohiro},
TITLE = {Thin-Shell Wormholes in Einstein and Einstein–Gauss–Bonnet Theories of Gravity},
JOURNAL = {Universe},
VOLUME = {6},
YEAR = {2020},
NUMBER = {11},
ARTICLE-NUMBER = {197},
URL = {https://www.mdpi.com/2218-1997/6/11/197},
ISSN = {2218-1997},
DOI = {10.3390/universe6110197}
}

@article{Fernandes_2022,
   title={The 4D Einstein–Gauss–Bonnet theory of gravity: a review},
   volume={39},
   ISSN={1361-6382},
   url={http://dx.doi.org/10.1088/1361-6382/ac500a},
   DOI={10.1088/1361-6382/ac500a},
   number={6},
   journal={Classical and Quantum Gravity},
   publisher={IOP Publishing},
   author={Fernandes, Pedro G S and Carrilho, Pedro and Clifton, Timothy and Mulryne, David J},
   year={2022},
   month=feb, pages={063001} }

@article{Odintsov_2020,
   title={Swampland implications of GW170817-compatible Einstein-Gauss-Bonnet gravity},
   volume={805},
   ISSN={0370-2693},
   url={http://dx.doi.org/10.1016/j.physletb.2020.135437},
   DOI={10.1016/j.physletb.2020.135437},
   journal={Physics Letters B},
   publisher={Elsevier BV},
   author={Odintsov, S.D. and Oikonomou, V.K.},
   year={2020},
   month=jun, pages={135437} }

@article{Kleihaus_2011,
   title={Rotating Black Holes in Dilatonic Einstein-Gauss-Bonnet Theory},
   volume={106},
   ISSN={1079-7114},
   url={http://dx.doi.org/10.1103/PhysRevLett.106.151104},
   DOI={10.1103/physrevlett.106.151104},
   number={15},
   journal={Physical Review Letters},
   publisher={American Physical Society (APS)},
   author={Kleihaus, Burkhard and Kunz, Jutta and Radu, Eugen},
   year={2011},
   month=apr }

@article{Silva_2018,
   title={Spontaneous Scalarization of Black Holes and Compact Stars from a Gauss-Bonnet Coupling},
   volume={120},
   ISSN={1079-7114},
   url={http://dx.doi.org/10.1103/PhysRevLett.120.131104},
   DOI={10.1103/physrevlett.120.131104},
   number={13},
   journal={Physical Review Letters},
   publisher={American Physical Society (APS)},
   author={Silva, Hector O. and Sakstein, Jeremy and Gualtieri, Leonardo and Sotiriou, Thomas P. and Berti, Emanuele},
   year={2018},
   month=mar }

@article{Doneva_2018,
   title={New Gauss-Bonnet Black Holes with Curvature-Induced Scalarization in Extended Scalar-Tensor Theories},
   volume={120},
   ISSN={1079-7114},
   url={http://dx.doi.org/10.1103/PhysRevLett.120.131103},
   DOI={10.1103/physrevlett.120.131103},
   number={13},
   journal={Physical Review Letters},
   publisher={American Physical Society (APS)},
   author={Doneva, Daniela D. and Yazadjiev, Stoytcho S.},
   year={2018},
   month=mar }

@article{Sotiriou_2014,
   title={Black Hole Hair in Generalized Scalar-Tensor Gravity},
   volume={112},
   ISSN={1079-7114},
   url={http://dx.doi.org/10.1103/PhysRevLett.112.251102},
   DOI={10.1103/physrevlett.112.251102},
   number={25},
   journal={Physical Review Letters},
   publisher={American Physical Society (APS)},
   author={Sotiriou, Thomas P. and Zhou, Shuang-Yong},
   year={2014},
   month=jun }

@misc{cai2025atomicquantumsensorshighfrequency,
      title={Atomic Quantum Sensors for High-Frequency Gravitational Wave Searches}, 
      author={Yi-Fu Cai and Luca Visinelli and Sheng-Feng Yan},
      year={2025},
      eprint={2510.15031},
      archivePrefix={arXiv},
      primaryClass={hep-ph},
      url={https://arxiv.org/abs/2510.15031}, 
}

@article{Figueroa_2023,
   title={CosmoLattice: A modern code for lattice simulations of scalar and gauge field dynamics in an expanding universe},
   volume={283},
   ISSN={0010-4655},
   url={http://dx.doi.org/10.1016/j.cpc.2022.108586},
   DOI={10.1016/j.cpc.2022.108586},
   journal={Computer Physics Communications},
   publisher={Elsevier BV},
   author={Figueroa, Daniel G. and Florio, Adrien and Torrenti, Francisco and Valkenburg, Wessel},
   year={2023},
   month=feb, pages={108586} }

@article{Figueroa_2021,
   title={The art of simulating the early universe.
Part I. Integration techniques and canonical cases},
   volume={2021},
   ISSN={1475-7516},
   url={http://dx.doi.org/10.1088/1475-7516/2021/04/035},
   DOI={10.1088/1475-7516/2021/04/035},
   number={04},
   journal={Journal of Cosmology and Astroparticle Physics},
   publisher={IOP Publishing},
   author={Figueroa, Daniel G. and Florio, Adrien and Torrenti, Francisco and Valkenburg, Wessel},
   year={2021},
   month=apr, pages={035} }

@misc{dankovsky2025cosmicdomainwallslattice,
      title={Cosmic domain walls on a lattice: illusive effects of initial conditions}, 
      author={I. Dankovsky and S. Ramazanov and E. Babichev and D. Gorbunov and A. Vikman},
      year={2025},
      eprint={2509.25367},
      archivePrefix={arXiv},
      primaryClass={hep-ph},
      url={https://arxiv.org/abs/2509.25367}, 
}

@article{Dankovsky_2024,
   title={Revisiting evolution of domain walls and their gravitational radiation with CosmoLattice},
   volume={2024},
   ISSN={1475-7516},
   url={http://dx.doi.org/10.1088/1475-7516/2024/09/047},
   DOI={10.1088/1475-7516/2024/09/047},
   number={09},
   journal={Journal of Cosmology and Astroparticle Physics},
   publisher={IOP Publishing},
   author={Dankovsky, I. and Babichev, E. and Gorbunov, D. and Ramazanov, S. and Vikman, A.},
   year={2024},
   month=sep, pages={047} }

@article{Abbott_2017,
    author = "Abbott, B. P. and others",
    collaboration = "LIGO Scientific, Virgo",
    title = "{GW170817: Observation of Gravitational Waves from a Binary Neutron Star Inspiral}",
    eprint = "1710.05832",
    archivePrefix = "arXiv",
    primaryClass = "gr-qc",
    reportNumber = "LIGO-P170817",
    doi = "10.1103/PhysRevLett.119.161101",
    journal = "Phys. Rev. Lett.",
    volume = "119",
    number = "16",
    pages = "161101",
    year = "2017"
}

@article{ZAHOOR2026100458,
title = {Reconciling fractional power potential and EGB gravity in the light of ACT},
journal = {Journal of High Energy Astrophysics},
volume = {49},
pages = {100458},
year = {2026},
month = jan,
issn = {2214-4048},
doi = {https://doi.org/10.1016/j.jheap.2025.100458},
url = {https://www.sciencedirect.com/science/article/pii/S2214404825001399},
author = {Mehnaz Zahoor and Suhail Khan and Imtiyaz Ahmad Bhat}
}

@article{Nojiri_2019,
   title={Ghost-free Gauss-Bonnet theories of gravity},
   volume={99},
   ISSN={2470-0029},
   url={http://dx.doi.org/10.1103/PhysRevD.99.044050},
   DOI={10.1103/physrevd.99.044050},
   number={4},
   journal={Physical Review D},
   publisher={American Physical Society (APS)},
   author={Nojiri, S. and Odintsov, S. D. and Oikonomou, V. K.},
   year={2019},
   month=feb }

@article{Pinto2025,
    author = {Miguel A. S. Pinto},
    title = {\uppercase{$\Lambda$CDM}-like evolution in Einstein-scalar-Gauss–Bonnet gravity},
    journal = {The European Physical Journal C},
    year = 2025,
    volume = 85,
    issn = {1434-6052},
    month = 9,
    doi = {https://doi.org/10.1140/epjc/s10052-025-14796-5}
}

@article{Abbott_2016,
    author = "Abbott, B. P. and others",
    collaboration = "LIGO Scientific, Virgo",
    title = "{Observation of Gravitational Waves from a Binary Black Hole Merger}",
    eprint = "1602.03837",
    archivePrefix = "arXiv",
    primaryClass = "gr-qc",
    reportNumber = "LIGO-P150914",
    doi = "10.1103/PhysRevLett.116.061102",
    journal = "Phys. Rev. Lett.",
    volume = "116",
    number = "6",
    pages = "061102",
    year = "2016"
}

@article{YOUSAF2025101221,
title = {Implications of modified Gauss–Bonnet gravity on gravastar-like structures: High-energy stability and electromagnetic effects},
journal = {High Energy Density Physics},
volume = {57},
pages = {101221},
year = {2025},
issn = {1574-1818},
doi = {https://doi.org/10.1016/j.hedp.2025.101221},
url = {https://www.sciencedirect.com/science/article/pii/S1574181825000497},
author = {M. Yousaf and H. Asad and Muhammad Aslam},
}

@misc{theligoscientificcollaboration2025upperlimitsisotropicgravitationalwave,
      title={Upper Limits on the Isotropic Gravitational-Wave Background from the first part of LIGO, Virgo, and KAGRA's fourth Observing Run}, 
      author={LIGO-Virgo-KAGRA},
      year={2025},
      eprint={2508.20721},
      archivePrefix={arXiv},
      primaryClass={gr-qc},
      url={https://arxiv.org/abs/2508.20721}, 
}

@article{Dolgov_2016,
   title={Evolution of thick domain walls in de Sitter universe},
   volume={2016},
   ISSN={1475-7516},
   url={http://dx.doi.org/10.1088/1475-7516/2016/10/026},
   DOI={10.1088/1475-7516/2016/10/026},
   number={10},
   journal={Journal of Cosmology and Astroparticle Physics},
   publisher={IOP Publishing},
   author={Dolgov, A.D. and Godunov, S.I. and Rudenko, A.S.},
   year={2016},
   month=oct, pages={026–026} }

@article{Dolgov_2018,
   title={Evolution of thick domain walls in inflationary and $p=w\rho $ universe},
   volume={78},
   ISSN={1434-6052},
   url={http://dx.doi.org/10.1140/epjc/s10052-018-6350-7},
   DOI={10.1140/epjc/s10052-018-6350-7},
   number={10},
   journal={The European Physical Journal C},
   publisher={Springer Science and Business Media LLC},
   author={Dolgov, A. D. and Godunov, S. I. and Rudenko, A. S.},
   year={2018},
   month=oct }

@article{Basu_1994,
   title={Evolution of topological defects during inflation},
   volume={50},
   ISSN={0556-2821},
   url={http://dx.doi.org/10.1103/PhysRevD.50.7150},
   DOI={10.1103/physrevd.50.7150},
   number={12},
   journal={Physical Review D},
   publisher={American Physical Society (APS)},
   author={Basu, Rama and Vilenkin, Alexander},
   year={1994},
   month=dec, pages={7150–7153} }

@article{PhysRevD.44.340,
  title = {Quantum creation of topological defects during inflation},
  author = {Basu, Rama and Guth, Alan H. and Vilenkin, Alexander},
  journal = {Phys. Rev. D},
  volume = {44},
  issue = {2},
  pages = {340--351},
  numpages = {0},
  year = {1991},
  month = Jul,
  publisher = {American Physical Society},
  doi = {10.1103/PhysRevD.44.340},
  url = {https://link.aps.org/doi/10.1103/PhysRevD.44.340}
}

@article{Dankovsky_2025,
doi = {10.1088/1475-7516/2025/02/064},
url = {https://dx.doi.org/10.1088/1475-7516/2025/02/064},
year = {2025},
month = feb,
publisher = {IOP Publishing},
volume = {2025},
number = {02},
pages = {064},
author = {Dankovsky, I. and Ramazanov, S. and Babichev, E. and Gorbunov, D. and Vikman, A.},
title = {Numerical analysis of melting domain walls and their gravitational waves},
journal = {Journal of Cosmology and Astroparticle Physics}
}

@article{Coleman_1973,
  title = {Radiative Corrections as the Origin of Spontaneous Symmetry Breaking},
  author = {Coleman, Sidney and Weinberg, Erick},
  journal = {Phys. Rev. D},
  volume = {7},
  issue = {6},
  pages = {1888--1910},
  numpages = {0},
  year = {1973},
  month = Mar,
  publisher = {American Physical Society},
  doi = {10.1103/PhysRevD.7.1888},
  url = {https://link.aps.org/doi/10.1103/PhysRevD.7.1888}
}

@article{Weinberg_1987,
  title = {Understanding complex perturbative effective potentials},
  author = {Weinberg, Erick J. and Wu, Aiqun},
  journal = {Phys. Rev. D},
  volume = {36},
  issue = {8},
  pages = {2474--2480},
  numpages = {0},
  year = {1987},
  month = Oct,
  publisher = {American Physical Society},
  doi = {10.1103/PhysRevD.36.2474},
  url = {https://link.aps.org/doi/10.1103/PhysRevD.36.2474}
}

@article{refId0,
	author = {{Planck Collaboration} and {Akrami, Y.} },
	title = {Planck 2018 results - X. Constraints on inflation},
	DOI= "10.1051/0004-6361/201833887",
	url= "https://doi.org/10.1051/0004-6361/201833887",
	journal = {Astronomy \& Astrophysics},
	year = 2020,
	volume = 641,
	pages = "A10",
}

@article{Sher_1989,
  title={Electroweak Higgs Potentials and Vacuum Stability},
  volume={179},
  ISSN={0370-1573},
  url={http://dx.doi.org/10.1016/0370-1573(89)90061-6},
  DOI={10.1016/0370-1573(89)90061-6},
  journal={Physics Reports},
  publisher={Elsevier},
  author={Sher, Marc},
  year={1989},
  month=apr,
  pages={273--418}
}

@article{Lyth_Stewart_1992,
  title={The curvature perturbation in power-law (e.g. extended) inflation},
  volume={46},
  ISSN={0556-2821},
  url={http://dx.doi.org/10.1103/PhysRevD.46.532},
  DOI={10.1103/PhysRevD.46.532},
  journal={Physical Review D},
  publisher={APS},
  author={Lyth, David H. and Stewart, Ewan D.},
  year={1992},
  month=jul,
  pages={532--538}
}

@article{Herranen_2014,
  title={Spacetime Curvature and the Higgs Stability During Inflation},
  volume={113},
  ISSN={0031-9007},
  url={http://dx.doi.org/10.1103/PhysRevLett.113.211102},
  DOI={10.1103/PhysRevLett.113.211102},
  journal={Physical Review Letters},
  publisher={APS},
  author={Herranen, Matti and Markkanen, Tommi and Nurmi, Sami and Rajantie, Arttu},
  year={2014},
  month=nov,
  pages={211102}
}

@article{Markkanen_2019,
  title={The decoupling limit in the presence of a cosmological constant},
  volume={793},
  ISSN={0370-2693},
  url={http://dx.doi.org/10.1016/j.physletb.2019.134830},
  DOI={10.1016/j.physletb.2019.134830},
  journal={Physics Letters B},
  publisher={Elsevier},
  author={Markkanen, Tommi and Tranberg, Anders},
  year={2019},
  month=nov,
  pages={134830}
}

@article{Guo_Schwarz_2010,
  title={Slow-roll inflation with a Gauss--Bonnet correction},
  volume={81},
  ISSN={1550-7998},
  url={http://dx.doi.org/10.1103/PhysRevD.81.123520},
  DOI={10.1103/PhysRevD.81.123520},
  journal={Physical Review D},
  publisher={APS},
  author={Guo, Zong-Kuan and Schwarz, Dominik J.},
  year={2010},
  month=jun,
  pages={123520}
}

@article{Kanti_2015,
  title={Gauss--Bonnet Inflation},
  volume={734},
  ISSN={0370-2693},
  url={http://dx.doi.org/10.1016/j.physletb.2014.11.013},
  DOI={10.1016/j.physletb.2014.11.013},
  journal={Physics Letters B},
  publisher={Elsevier},
  author={Kanti, Panagiota and Gannouji, Radouane and Dadhich, Naresh},
  year={2015},
  month=jan,
  pages={27--33}
}

@article{Weinberg_2005,
  title={Quantum contributions to cosmological correlations},
  volume={72},
  ISSN={1550-7998},
  url={http://dx.doi.org/10.1103/PhysRevD.72.043514},
  DOI={10.1103/PhysRevD.72.043514},
  journal={Physical Review D},
  publisher={APS},
  author={Weinberg, Steven},
  year={2005},
  month=aug,
  pages={043514}
}

@article{Parker_Toms_1985,
  title={New form for the coincidence limit of the Feynman propagator, or heat kernel, in curved spacetime},
  volume={31},
  ISSN={0556-2821},
  url={http://dx.doi.org/10.1103/PhysRevD.31.953},
  DOI={10.1103/PhysRevD.31.953},
  journal={Physical Review D},
  publisher={APS},
  author={Parker, Leonard and Toms, David J.},
  year={1985},
  month=feb,
  pages={953--957}
}

@article{Liang:2019fkj,
    author = "Liang, Qiuyue and Sakstein, Jeremy and Trodden, Mark",
    title = "{Baryogenesis via gravitational spontaneous symmetry breaking}",
    eprint = "1904.10510",
    archivePrefix = "arXiv",
    primaryClass = "hep-ph",
    doi = "10.1103/PhysRevD.100.063518",
    journal = "Phys. Rev. D",
    volume = "100",
    number = "6",
    pages = "063518",
    year = "2019"
}

@article{Kanti:2015pda,
    author = "Kanti, Panagiota and Gannouji, Radouane and Dadhich, Naresh",
    title = "{Gauss-Bonnet Inflation}",
    eprint = "1503.01579",
    archivePrefix = "arXiv",
    primaryClass = "hep-th",
    doi = "10.1103/PhysRevD.92.041302",
    journal = "Phys. Rev. D",
    volume = "92",
    number = "4",
    pages = "041302",
    year = "2015"
}

@article{Jiang:2013gza,
    author = "Jiang, Peng-Xu and Hu, Jian-Wei and Guo, Zong-Kuan",
    title = "{Inflation coupled to a Gauss-Bonnet term}",
    eprint = "1310.5579",
    archivePrefix = "arXiv",
    primaryClass = "hep-th",
    doi = "10.1103/PhysRevD.88.123508",
    journal = "Phys. Rev. D",
    volume = "88",
    pages = "123508",
    year = "2013"
}

@article{Hwang:1999gf,
    author = "Hwang, Jai-chan and Noh, Hyerim",
    title = "{Conserved cosmological structures in the one loop superstring effective action}",
    eprint = "astro-ph/9909480",
    archivePrefix = "arXiv",
    doi = "10.1103/PhysRevD.61.043511",
    journal = "Phys. Rev. D",
    volume = "61",
    pages = "043511",
    year = "2000"
}

@article{Cartier:2001is,
    author = "Cartier, Cyril and Hwang, Jai-chan and Copeland, Edmund J.",
    title = "{Evolution of cosmological perturbations in nonsingular string cosmologies}",
    eprint = "astro-ph/0106197",
    archivePrefix = "arXiv",
    reportNumber = "SUSX-TH-01-026",
    doi = "10.1103/PhysRevD.64.103504",
    journal = "Phys. Rev. D",
    volume = "64",
    pages = "103504",
    year = "2001"
}

@article{Hwang:2005hb,
    author = "Hwang, Jai-chan and Noh, Hyerim",
    title = "{Classical evolution and quantum generation in generalized gravity theories including string corrections and tachyon: Unified analyses}",
    eprint = "gr-qc/0412126",
    archivePrefix = "arXiv",
    doi = "10.1103/PhysRevD.71.063536",
    journal = "Phys. Rev. D",
    volume = "71",
    pages = "063536",
    year = "2005"
}

@article{Guo:2009uk,
    author = "Guo, Zong-Kuan and Schwarz, Dominik J.",
    title = "{Power spectra from an inflaton coupled to the Gauss-Bonnet term}",
    eprint = "0907.0427",
    archivePrefix = "arXiv",
    primaryClass = "hep-th",
    reportNumber = "BI-TP-2009-15",
    doi = "10.1103/PhysRevD.80.063523",
    journal = "Phys. Rev. D",
    volume = "80",
    pages = "063523",
    year = "2009"
}

@article{Guo:2010jr,
    author = "Guo, Zong-Kuan and Schwarz, Dominik J.",
    title = "{Slow-roll inflation with a Gauss-Bonnet correction}",
    eprint = "1001.1897",
    archivePrefix = "arXiv",
    primaryClass = "hep-th",
    reportNumber = "BI-TP-2010-01",
    doi = "10.1103/PhysRevD.81.123520",
    journal = "Phys. Rev. D",
    volume = "81",
    pages = "123520",
    year = "2010"
}

@article{Koh:2014bka,
    author = "Koh, Seoktae and Lee, Bum-Hoon and Lee, Wonwoo and Tumurtushaa, Gansukh",
    title = "{Observational constraints on slow-roll inflation coupled to a Gauss-Bonnet term}",
    eprint = "1404.6096",
    archivePrefix = "arXiv",
    primaryClass = "gr-qc",
    doi = "10.1103/PhysRevD.90.063527",
    journal = "Phys. Rev. D",
    volume = "90",
    number = "6",
    pages = "063527",
    year = "2014"
}

@article{Kawai:1998ab,
    author = "Kawai, Shinsuke and Sakagami, Masa-aki and Soda, Jiro",
    title = "{Instability of one loop superstring cosmology}",
    eprint = "gr-qc/9802033",
    archivePrefix = "arXiv",
    reportNumber = "KUCP-0114",
    doi = "10.1016/S0370-2693(98)00925-3",
    journal = "Phys. Lett. B",
    volume = "437",
    pages = "284--290",
    year = "1998"
}

@article{Kawai:2017kqt,
    author = "Kawai, Shinsuke and Kim, Jinsu",
    title = "{Gauss\textendash{}Bonnet Chern\textendash{}Simons gravitational wave leptogenesis}",
    eprint = "1702.07689",
    archivePrefix = "arXiv",
    primaryClass = "hep-th",
    doi = "10.1016/j.physletb.2018.12.019",
    journal = "Phys. Lett. B",
    volume = "789",
    pages = "145--149",
    year = "2019"
}

@article{Yi:2018gse,
    author = "Yi, Zhu and Gong, Yungui and Sabir, Mudassar",
    title = "{Inflation with Gauss-Bonnet coupling}",
    eprint = "1804.09116",
    archivePrefix = "arXiv",
    primaryClass = "gr-qc",
    doi = "10.1103/PhysRevD.98.083521",
    journal = "Phys. Rev. D",
    volume = "98",
    number = "8",
    pages = "083521",
    year = "2018"
}

@article{Rashidi:2020wwg,
    author = "Rashidi, Narges and Nozari, Kourosh",
    title = "{Gauss-Bonnet Inflation after Planck2018}",
    eprint = "2001.07012",
    archivePrefix = "arXiv",
    primaryClass = "astro-ph.CO",
    doi = "10.3847/1538-4357/ab6a10",
    journal = "Astrophys. J.",
    volume = "890",
    pages = "58",
    month = "1",
    year = "2020"
}

@article{Kawai:2021bye,
    author = "Kawai, Shinsuke and Kim, Jinsu",
    title = "{CMB from a Gauss-Bonnet-induced de Sitter fixed point}",
    eprint = "2105.04386",
    archivePrefix = "arXiv",
    primaryClass = "hep-ph",
    reportNumber = "CERN-TH-2021-075",
    doi = "10.1103/PhysRevD.104.043525",
    journal = "Phys. Rev. D",
    volume = "104",
    number = "4",
    pages = "043525",
    year = "2021"
}

@article{Kawai:2021edk,
    author = "Kawai, Shinsuke and Kim, Jinsu",
    title = "{Primordial black holes from Gauss-Bonnet-corrected single field inflation}",
    eprint = "2108.01340",
    archivePrefix = "arXiv",
    primaryClass = "astro-ph.CO",
    reportNumber = "CERN-TH-2021-115",
    doi = "10.1103/PhysRevD.104.083545",
    journal = "Phys. Rev. D",
    volume = "104",
    number = "8",
    pages = "083545",
    year = "2021"
}

@article{Kawaguchi:2022nku,
    author = "Kawaguchi, Ryodai and Tsujikawa, Shinji",
    title = "{Primordial black holes from Higgs inflation with a Gauss-Bonnet coupling}",
    eprint = "2211.13364",
    archivePrefix = "arXiv",
    primaryClass = "astro-ph.CO",
    reportNumber = "WUCG-22-11",
    doi = "10.1103/PhysRevD.107.063508",
    journal = "Phys. Rev. D",
    volume = "107",
    number = "6",
    pages = "063508",
    year = "2023"
}

@article{Addazi:2024gew,
    author = "Addazi, Andrea and Aldabergenov, Yermek and Cai, Yifu",
    title = "{Sound speed resonance of gravitational waves in Gauss-Bonnet-coupled inflation}",
    eprint = "2408.05091",
    archivePrefix = "arXiv",
    primaryClass = "gr-qc",
    doi = "10.1103/PhysRevD.110.123530",
    journal = "Phys. Rev. D",
    volume = "110",
    number = "12",
    pages = "123530",
    year = "2024"
}

@article{Hindmarsh:2013xza,
    author = "Hindmarsh, Mark and Huber, Stephan J. and Rummukainen, Kari and Weir, David J.",
    title = "{Gravitational waves from the sound of a first order phase transition}",
    eprint = "1304.2433",
    archivePrefix = "arXiv",
    primaryClass = "hep-ph",
    reportNumber = "HIP-2013-07-TH",
    doi = "10.1103/PhysRevLett.112.041301",
    journal = "Phys. Rev. Lett.",
    volume = "112",
    pages = "041301",
    year = "2014"
}

@article{Addazi:2023jvg,
    author = "Addazi, Andrea and Cai, Yi-Fu and Marciano, Antonino and Visinelli, Luca",
    title = "{Have pulsar timing array methods detected a cosmological phase transition?}",
    eprint = "2306.17205",
    archivePrefix = "arXiv",
    primaryClass = "astro-ph.CO",
    reportNumber = "CA21106; CA21136",
    doi = "10.1103/PhysRevD.109.015028",
    journal = "Phys. Rev. D",
    volume = "109",
    number = "1",
    pages = "015028",
    year = "2024"
}

@article{Addazi:2017nmg,
    author = "Addazi, Andrea and Cai, Yi-Fu and Marciano, Antonino",
    title = "{Testing Dark Matter Models with Radio Telescopes in light of Gravitational Wave Astronomy}",
    eprint = "1712.03798",
    archivePrefix = "arXiv",
    primaryClass = "hep-ph",
    doi = "10.1016/j.physletb.2018.06.015",
    journal = "Phys. Lett. B",
    volume = "782",
    pages = "732--736",
    year = "2018"
}

@article{Addazi:2020zcj,
    author = "Addazi, Andrea and Cai, Yi-Fu and Gan, Qingyu and Marciano, Antonino and Zeng, Kaiqiang",
    title = "{NANOGrav results and dark first order phase transitions}",
    eprint = "2009.10327",
    archivePrefix = "arXiv",
    primaryClass = "hep-ph",
    doi = "10.1007/s11433-021-1724-6",
    journal = "Sci. China Phys. Mech. Astron.",
    volume = "64",
    number = "9",
    pages = "290411",
    year = "2021"
}

@article{Addazi:2024osi,
    author = "Addazi, Andrea and Capozziello, Salvatore and Gan, Qingyu",
    title = "{Resonant graviton-photon transitions with cosmological stochastic magnetic field}",
    doi = "10.1016/j.physletb.2024.138574",
    journal = "Phys. Lett. B",
    volume = "851",
    pages = "138574",
    year = "2024"
}

@article{Addazi:2024kbq,
    author = "Addazi, Andrea and Capozziello, Salvatore and Gan, Qingyu",
    title = "{Resonant graviton{\textendash}photon conversion with stochastic magnetic field in the expanding universe}",
    eprint = "2401.15965",
    archivePrefix = "arXiv",
    primaryClass = "gr-qc",
    doi = "10.1016/j.dark.2025.101844",
    journal = "Phys. Dark Univ.",
    volume = "48",
    pages = "101844",
    year = "2025"
}

\end{document}